\documentclass[aps,amssymb,prb, twocolumn]{revtex4}
\usepackage{graphicx}
\usepackage{epsfig,amsmath}

\begin{document}

%\title{Transition effects induced by frequency-dependent third cumulant of current fluctuations on a small quantum system}
\title{Global relaxation in superconducting qubits}

\author{T.~Ojanen$^1$}
\email[Correspondence to ]{teemuo@boojum.hut.fi}
\author{A. O. Niskanen$^{2,3}$}
\author{Y. Nakamura$^{2,4,5}$}
\author{A. A. Abdumalikov Jr.$^{5,6}$}
 \affiliation{ $^1$Low Temperature Laboratory, Helsinki University of
Technology, P.~O.~Box 2200, FIN-02015 HUT, Finland }
 \affiliation{$^2$ CREST-JST, Kawaguchi, Saitama 332-0012, Japan }
\affiliation{$^3$ VTT Technical Research Centre of Finland, Sensors,
P.O. Box 1000, 02044 VTT, Finland} \affiliation{$^4$ NEC Fundamental
and Environmental Research Laboratories, Tsukuba, Ibaraki 305-8501,
Japan} \affiliation{$^5$ The Institute of Physical and Chemical
Research (RIKEN), Wako, Saitama 351-0198, Japan}
\affiliation{$^6$Physical-Technical Institute of the Academy of
Sciences, Mavlyanov-str. 2B, Tashkent, 700084, Uzbekistan }
\date{\today}
\begin{abstract}

We consider coupled quantum two-state systems (qubits) exposed to a
global relaxation process. The global relaxation refers to the
assumption that qubits are coupled to the \emph{same} quantum bath
with approximately equal strengths, appropriate for long-wavelength
environmental fluctuations. We show that interactions do not spoil
the picture of Dicke's subradiant and superradiant states where
quantum interference effects lead to striking deviations from the
independent relaxation picture. Remarkably, the system possess a
stable entangled state and a state decaying faster than single qubit
excitations. We propose a scheme how these effects can be
experimentally accessed in superconducting flux qubits and,
possibly, used in constructing long-lived entangled states.

\end{abstract}
\pacs{PACS numbers: } \bigskip

\maketitle

A lot of experimental progress has been made in superconducting
qubits recently including the achievement of several $\mu$s
coherence times.\cite{saclay,bertet,yoshihara,wallraffprl} High
visibilities\cite{wallraffprl,steffenprl,siddiqiprb} and even
nondemolition\cite{lupascu} readout has been demonstrated. Several
coupled-qubit experiments have been also carried out, see e.g. Refs.
\onlinecite{yamamoto,steffen,hime,myscience}. However, energy
relaxation has proved to be a serious limitation to the coherence in
quantum information applications. The origin and the detailed
mechanism of relaxation has remained largely unknown.
%While single qubit decoherence has been studied
%theoretically \cite{something} and experimentally \cite{more} in detail, the decoherence in
%multi-qubit systems has not received as much attention.

We consider a two qubit system, where qubits feel the same
fluctuating quantum bath. We concentrate on an interacting
generalization of the well-known Dicke model\cite{dicke}, which is
relevant in the case of long-wavelength spontaneous emission induced
by the environment. Dicke studied a spontaneous emission of an
ensemble of non-interacting molecules coupled to a common bath and
predicted large deviations from the independent relaxation picture.
He showed that certain correlated states decay more rapidly
(superradiance) or are more stable than uncorrelated excitations
(subradiance). The existence of subradiant and superradiant states
was decisively observed much later in spontaneous emission of two
nearby trapped atoms.\cite{devoe} Since then, correlated decay of
states has been studied experimentally and theoretically in quantum
dot and double dot systems where the Dicke-type behavior has been
observed.\cite{fujisawa,brandes,vorrath,stolcz1,stolcz2} Recently it
was discovered that the subradiant states can be employed in
optimizing multi-qubit quantum algorithms in the presence of global
relaxation.\cite{utsunomiya}

In this Letter we demonstrate how the different energy states of
interacting qubits may decay very differently under global
relaxation due to quantum interference effects. As in the case of
noninteracting molecules, there exist a stable entangled state and a
state that decays faster than uncorrelated excited state. Testing
the validity of the correlated decay in the context of
superconducting flux qubits is discussed in detail. By studying the
decay of different two-qubit states, one can obtain information of
the presently unknown relaxation mechanism that is inaccessible in
single qubit experiments. Currently it is not understood whether the
limiting intrinsic relaxation is caused by high-frequency flux noise
or something else. See e.g.
Refs.~\onlinecite{bertet,yoshihara,kakuyanagi} for some experimental
data.

We consider a system consisting of qubits coupled to a
relaxation-inducing quantum bath described by the Hamiltonian
$H=H_q+H_{\mathrm{env}}+H_{\mathrm{i}}$, where
\begin{align}\label{eka}
H_q=-\frac{\Delta}{2}\sum_i\sigma_z^{(i)}+J\sum_{i<j}
\sigma_x^{(i)}\sigma_x^{(j)},\,
H_{\mathrm{i}}=g\hat{x}\sum_i\sigma_x^{(i)}.
\end{align}
Here $\hat{x}$ is a Hermitian operator in the environment part of
the Hilbert space. The many-body Hamiltonian of the environment
$H_{\mathrm{env}}$ does not need to be specified in detail, its
effects enter through correlation functions of $\hat{x}$. It is
assumed that qubits have equal energy splittings $\Delta$,
interaction strengths $J$ and bath coupling constants $g$. These
features are realized in the case of similar qubits in the close
proximity compared to the relevant length scale of environment
fluctuations. Below we estimate effects due to detuning of
parameters. The form of the coupling on Eq.~(\ref{eka}) is assumed
to be $\sigma_x\otimes\sigma_x$-type as this is natural for
so-called optimally biased superconducting qubits as will become
apparent below. Also the $\sigma_x$ type coupling to the environment
is natural since the effect of longitudinal coupling is strongly
suppressed. Moreover, we are focusing
on the effect of relaxation which is not affected by longitudinal coupling. %sigma_y coupling?

The evolution of the total system obeys the von Neumann equation
$\dot{\rho}_T(t)=-\frac{i}{\hbar}[H,\rho_T(t)]$, which is formally
solved by $\rho_T(t)=U(t,t_0)\rho_T^0U^{\dagger}(t,t_0)$, where
$U(t,t_0)=\mathrm{exp}(-iH(t-t_0)/\hbar)$. For a factorizable
initial state $\rho_T^0=\rho^0\otimes\rho_{\mathrm{env}}^0$, the
reduced density matrix for qubits can be written in terms of a
propagator by $\rho_{ij}(t)=G(ij,t;kl,t_0)\rho_{kl}^0$ (summation
over repeated indices), where
\begin{align}\label{prop}
G(ij,t;kl,t_0)=&\mathrm{Tr}_{\mathrm{env}}\left[
\rho_{\mathrm{env}}^0 \langle l, t_0| \widetilde
{T}e^{\frac{i}{\hbar}\int_{t'}^{t}
d\bar{t}H_{\mathrm{int}}(\bar{t})} |j,t\rangle \right.\times\nonumber\\
&\left.\times\langle i, t| Te^{-\frac{i}{\hbar}\int_{t'}^{t}
d\bar{t}H_{\mathrm{int}}(\bar{t})} |k,t_0\rangle \right].
\end{align}
Expression (\ref{prop}) is written in the interaction picture, where
$H_{\mathrm{int}}(t)=
e^{\frac{i}{\hbar}(H_q+H_{\mathrm{env}})t}H_{\mathrm{int}}e^{-\frac{i}{\hbar}(H_q+H_{\mathrm{env}})t}
$ and $T$, $\widetilde {T}$ are time and antitime ordering
operators. In the case of two qubits the relevant Hilbert space is
spanned by the vectors
$|--\rangle\equiv|1\rangle,|-+\rangle\equiv|2\rangle,|+-\rangle\equiv|3\rangle,|++\rangle\equiv|4\rangle$.
First we study the case $J=0$, so the basis vectors are also
eigenstates of $H_q$.  Supposing that the environment is at low
temperature, excited states decay to the groundstate $|1\rangle$.
The transition rate $\Gamma_{\rho^0\to 1}$, defined as the linearly
growing contribution to the probability $\rho_{11}$ in the long-time
evolution, can be calculated from Eq. (\ref{prop}) by
$\Gamma_{\rho^0\to 1}=\lim_{T\to \infty}
\frac{G(11,T/2;ij,-T/2)}{T}\rho_{ij}^0$, where $\rho^0$ corresponds
to a stationary state. Expanding the propagator to the lowest
non-vanishing order, one recovers the Golden-Rule results
\begin{equation}\label{reitti1}
\Gamma_{2\to 1}=\Gamma_{3\to
1}=\frac{g^2}{\hbar^2}S_x(\Delta/\hbar),
\end{equation}
where
$S_x(\omega)=\int_{-\infty}^{\infty}\langle\hat{x}(t)\hat{x}(0)\rangle
e^{i\omega t}dt$. The transition rates are proportional to the noise
power at frequency $\Delta/\hbar$. These results are structurally
similar to the ones obtained in the case of independent baths for
each qubit. Interference effects come into play in the decay of the
correlated excited states
$|\phi_s\rangle=(|+-\rangle+|-+\rangle)/\sqrt{2}$ and
$|\phi_a\rangle=(|+-\rangle-|-+\rangle)/\sqrt{2}$. By performing an
analogous calculation we obtain $\Gamma_{\phi_{s}\to
1}=2\Gamma_{2\to 1}$. The rate enhancement is a direct evidence of
the global nature of the relaxation process. Interference effects
have even more dramatic impact on the evolution of $|\phi_a\rangle$
since it does not decay at all. This statement does not rely on the
perturbation theory and is an exact consequence of the dynamics
generated by (\ref{eka}). This is in a striking contrast to the case
where the two qubits are exposed to independent environment
fluctuations. In the case of finite interaction $J\neq0$, the above
described picture remains qualitatively the same. Now the system has
four non-degenerate eigenstates $|d\rangle\equiv
a|1\rangle+b|4\rangle$, $|\phi_s\rangle$, $|\phi_a\rangle$ and
$|u\rangle\equiv-b|1\rangle+a|4\rangle$ with respective energies
$-\sqrt{\Delta^2+J^2}$, $J$, -$J$ and $\sqrt{\Delta^2+J^2}$. The
coefficients are given by
$a=(1+\Delta/(2\sqrt{J^2+\Delta^2}))^{1/2}$ and
$b=-(1-\Delta/(2\sqrt{J^2+\Delta^2}))^{1/2}$. The decay rate of the
symmetric excitation is
\begin{equation}\label{hajonta}
\Gamma_{\phi_s\to
d}=\frac{g^2}{\hbar^2}2(a+b)^2S_x(\frac{\sqrt{\Delta^2+J^2}+J}{\hbar}),
\end{equation}
while $|\phi_a\rangle$ still remains exactly stable.

%Thus, depending on the initial state, the qubits may interfere
%constructively or destructively giving rise to large deviations from
%the independent relaxation picture. The described phenomenon could
%be be used as a tool in probing the nature of the relaxation and
%extracting information of the underlying physical processes. For
%example, if fluctuations contributing to the relaxation in
%superconducting qubits have much longer wavelength than the sample
%size, the decay of the entangled two qubit states should show
%increase or decrease compared to uncorrelated states.

Contrary to what was assumed in Eq.~(\ref{eka}), the bath couplings
of qubits never coincide exactly in experimental realizations. Also
when qubits are realized artificially, for example, by quantum dots
or superconducting circuits, individual Hamiltonians are not
identical but depend on material parameters and sample-specific
geometries. These features lead to deviations from the model
(\ref{eka}) and modifies previous conclusions to some extent.
Assuming the qubits are coupled to the bath with couplings
 $g_1, g_2$, the relaxation rates for $\phi_j$ ($j=s,a$)
become
\begin{align}\label{dev}
&\Gamma_{\phi_j\to d}=\frac{(g_1\pm g_2)^2}{2\hbar^2}(a+b)^2
S_x(\frac{\sqrt{\Delta^2+J^2}\pm J}{\hbar}),
%&\Gamma_{\phi_a\to d}=\frac{(g_1-g_2)^2}{2\hbar^2}(a+b)^2
%S_x(\frac{\sqrt{\Delta^2+J^2}-J}{\hbar}).
\end{align}
where upper signs correspond to $j=s$. The decay of subradiant state
$|\phi_a\rangle$ vanishes as a square of the detuning $g_1-g_2$. In
the case $J=0$ or when the noise is fairly insensitive to variations
of magnitude $J$ around $\sqrt{\Delta^2+J^2}$, the decay rates are
related by $\Gamma_{\phi_s\to 1}/\Gamma_{\phi_a\to
1}=(g_1+g_2)^2/(g_1-g_2)^2$,
%\begin{equation}\label{fraction}
%\frac{\Gamma_{\phi_s\to 1}}{\Gamma_{\phi_a\to 1}}=
%\frac{(g_1+g_2)^2}{(g_1-g_2)^2},
%\end{equation}
clearly demonstrating a dramatic difference when $g_1\approx g_2$.
Thus $|\phi_a\rangle$ is robust against fluctuations and
$|\phi_s\rangle$ decays rapidly even when the bath couplings match
only approximately. Let's assume now that the qubits have slightly
different energies $\Delta_1$ and $\Delta_2$. To simplify following
expressions we define functions
\begin{align}\label{lyhennys}
&a(x)=\frac{1}{\sqrt{2}}\left(1+\frac{x}{\sqrt{x^2+J^2}}\right)^{\frac{1}{2}},\\
&b(x)=-\frac{1}{\sqrt{2}}\left(1-\frac{x}{\sqrt{x^2+J^2}}\right)^{\frac{1}{2}}.
\end{align}
The eigenstates become $|\tilde{d}\rangle\equiv
a(\Delta_s)|1\rangle+b(\Delta_s)|4\rangle$,
$|\tilde{\phi}_s\rangle\equiv
-b(\Delta_a)|2\rangle+a(\Delta_a)|3\rangle$ ,
$|\tilde{\phi}_a\rangle\equiv
a(\Delta_a)|2\rangle+b(\Delta_a)|3\rangle$ and
$|\tilde{u}\rangle\equiv-b(\Delta_s)|1\rangle+a(\Delta_s)|4\rangle$,
where $\Delta_s=(\Delta_1+\Delta_2)/2$ and
$\Delta_a=(\Delta_1-\Delta_2)/2$. The states have respective
energies $E_{\tilde{d}/\tilde{u}}=\mp\sqrt{\Delta_s^2+J^2}$,
$E_{\tilde{\phi}_s/\tilde{\phi}_a}=\pm\sqrt{\Delta_a^2+J^2}$. In the
limit of vanishing bath coupling detuning the rates become
\begin{align}\label{edetuning}
&\Gamma_{\tilde{\phi}_j\to \tilde{d}}=\frac{g^2}{\hbar^2}
(a(\Delta_s)+b(\Delta_s))^2\times \nonumber\\
&\quad\times(a(\Delta_a)\mp b(\Delta_a))^2
S_x(\frac{E_{\tilde{\phi}_j}-E_{\tilde{d}}}{\hbar}),
%&\Gamma_{\tilde{\phi}_a\to \tilde{d}}=
%\frac{g^2}{\hbar^2}(a(\Delta_s)+b(\Delta_s))^2
%\times\nonumber\\
%&\quad\times(a(\Delta_a)+b(\Delta_a))^2
%S_x(\frac{E_{\tilde{\phi}_a}-E_{\tilde{d}}}{\hbar}).
\end{align}
where the minus sign corresponds to $j=s$. In the regime
$|\Delta_a|/J\ll 1$ these expressions can be estimated by
\begin{align}\label{edetuning2}
&\Gamma_{\tilde{\phi}_s\to \tilde{d}}=\frac{2g^2}{\hbar^2}
(a(\Delta_s)+b(\Delta_s))^2S_x(\frac{E_{\tilde{\phi}_s}-E_{\tilde{d}}}{\hbar}),\nonumber\\
&\Gamma_{\tilde{\phi}_a\to \tilde{d}}=
\frac{2g^2}{\hbar^2}(a(\Delta_s)+b(\Delta_s))^2
\left(\frac{\Delta_a}{2J}\right)^2
S_x(\frac{E_{\tilde{\phi}_a}-E_{\tilde{d}}}{\hbar}),\nonumber
\end{align}
implying that $|\tilde{\phi}_a\rangle$ maintains its subradiant
nature when detuning is small compared to the inter-qubit coupling.

\begin{figure*}
\begin{picture}(150,150)
\put(-160,0){\includegraphics[width=0.95\textwidth]{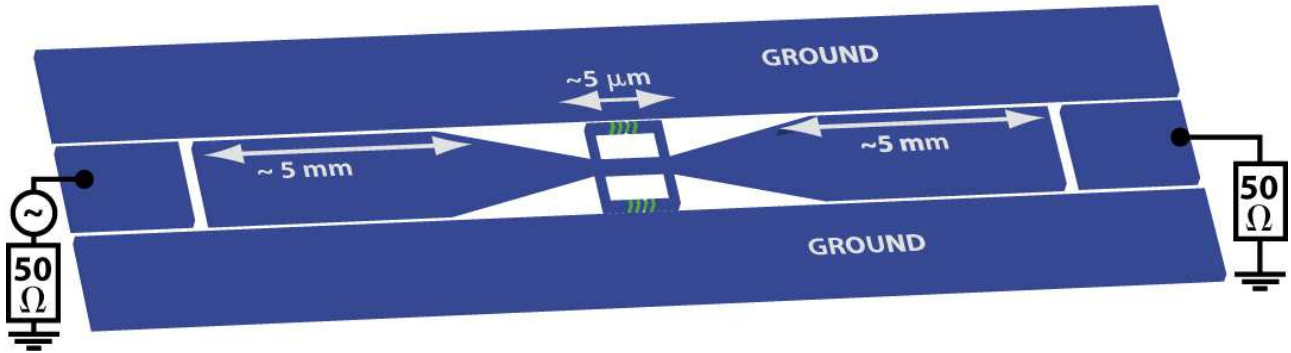}}
\end{picture}
\caption{\label{fg:scheme} Schematic of the suggested experiment.
The dimensions are exaggerated for clarity. The disconnected section
in the middle forms a coplanar resonator whose resonant frequency is
modified depending on the qubit state thus allowing for dispersive
readout. The chirality is such that the control microwave input via
the same port as the readout couples antisymmetrically to the
qubit.}
\end{figure*}

To study the nature of the relaxation process we suggest a system of
two flux qubits\cite{mooij,chiorescu} with as indentical parameters
as possible coupled to a high-Q cavity\cite{wallraff}, see Fig.~1.
We will now discuss a numerical example to show that the phenomenon
is indeed very spectacular even in the presence of imperfections
provided the assumption of globality of the noise holds. As shown
above, using a large coupling energy protects against any parameter
fluctuations and therefore the assumption of identical qubits is
quite realistic. The qubit $j$ ($j$=1,2) subspace when biased at the
half-flux quantum point $\Phi_0/2$ consists of two circulating
current states carrying a current of $\pm I_{\rm p}^j$. Tunneling
between the states happens at a rate of $\Delta_j/\hbar$. Neglecting
the off-resonant coupling to the cavity (used for dispersive
readout), the qubits are described explicitly by the Hamiltonian
\begin{equation}\label{expe}
H_q=-\sum_{j=1}^{2}\left(\frac{\Delta_j}{2}\sigma_z^{(j)}-\frac{\epsilon_j}{2}\sigma_x^{(j)}\right)
+J\sigma_x^{(1)}\sigma_x^{(2)}
\end{equation}
At the optimal point $\varepsilon_j=2I_{\rm p}^j(\Phi-\Phi_0/2)=0$
dephasing due to low-frequency flux fluctuations is minimized. To
achieve symmetry and to optimize coherence we assume
$\varepsilon_1\approx\varepsilon_2\approx 0$ and
$\Delta_1\approx\Delta_2$. As shown above $|\Delta_2-\Delta_1|$
should be compared to $J=MI_{\rm p}^1I_{\rm p}^2$ where $M$ is the
mutual (kinetic) inductance between the qubit loops. A realistic
sample\cite{hime,jaenis} may have quite similar tunneling energies
and a large coupling so as an example we assume
$(\Delta_2-\Delta_1)/h=200$ MHz, $\Delta_1/h=6$ GHz and $J/h=1$ GHz.
Choosing the bias of one of the qubits, say qubit 2, to be
$\varepsilon_2=0$ is easy using a global magnetic field and a
typical e-beam patterned sample with nominally same area may then
have $\varepsilon_1/h=200$ MHz. This last assumption further
modifies the eigenstates $|\tilde{d}\rangle$,
$|\tilde{\phi}_s\rangle$, $|\tilde{\phi}_a\rangle$ and
$|\tilde{u}\rangle$. These are reasonable and quite conservative
assumptions as the suggested sample geometry has perfect symmetry
about the center conductor and e-beam patterning is very accurate. A
numerical calculation then gives for a symmetric coupling energy $g$
% just using FGR and matrix elements
\begin{align}\label{example}
&\Gamma_{\tilde{\phi}_s\to \tilde{d}}=1.7\times\frac{g^2}{\hbar^2} S_x(2\pi\times 7.2\,{\rm GHz}),\\
&\Gamma_{\tilde{\phi}_a\to \tilde{d}}=4.0\times
10^{-3}\times\frac{g^2}{\hbar^2} S_x(2\pi\times 5.2\,{\rm GHz}).
\end{align}
Assuming that the noise spectrum $S_x (\omega)$ does  not have too
strong frequency dependence we then expect two orders of magnitude
different relaxation times for the sub- and superradiant states even
with very typical parameters. As shown in the beginning of the
paper, the factor $g^2/\hbar^2 S_x (\omega)$ appearing in the above
formulas is the characteristic relaxation rate for individual
qubits. This could be typically, say, 1 $\mu$s. This translates into
a 250 $\mu$s lifetime of the antisymmetric state under global noise
while the symmetric state decays in about 0.6 $\mu$s. Considering
that presently energy relaxation is limiting coherence in our flux
qubits\cite{yoshihara} very long overall coherence can then be
expected if a significant amount of the high-frequency noise is
global. The large coupling energy $J$ not only protects from
parameter scatter but also provides a gap of about $2J$ between
$|\tilde{\phi}_s\rangle$ and $|\tilde{\phi}_a\rangle$. Although this
transition is suppressed for single-qubit noise (flipping both
qubits required), it is better to have the difference as large as
possible to avoid stimulated emission.

The apparent contradiction in the present setting is on one hand the
stability of $|\phi_a\rangle$ under any kind of global
high-frequency field and on the other hand the desire to excite the
transition. It is clear that a symmetric drive cannot achieve this,
as demonstrated in Ref.~\onlinecite{hime}. As shown schematically in
Fig.~\ref{fg:scheme} we therefore assume that the qubits are coupled
anti-symmetrically (due to the left- and right-handed configurations
of the qubits) to the center conductor such that a resonant drive
via the transmission line can excite the $|d\rangle
\leftrightarrow|\phi_a\rangle$ transition and ideally only that.
That is, the microwave Hamiltonian can be estimated as $H_{\rm
mw}=\alpha(t)(\sigma_x^{(2)}-\sigma_x^{(1)})$ (if the drive and
cavity are far detuned from the cavity angular frequency $\omega$)
for which clearly the excitation of $|\phi_a\rangle$ is possible
since $\langle d|(\sigma_x^{(2)}-\sigma_x^{(1)})|\phi_a\rangle \neq
0$ but transitions between the symmetric states are forbidden. The
anti-symmetric microwave drive amplitude $\alpha(t)$ obeys
$\alpha(t)=\delta\Phi(t)I_{\rm p}$ where $I_{\rm p}$ is the
persistent current of the qubit and $\delta\Phi(t)$ is the ac flux
drive.

The coupling to the transmission line cavity has to be weak enough
such that the anti-symmetric coupling does not allow for significant
relaxation to the 50 $\Omega$ environment due to the finite quality
factor $Q$ of the cavity. In the case of a transmission measurement
and coupling via current it is most natural to use a half wavelength
resonance since this mode has an antinode of voltage and a node of
current in the middle. Also all other modes are guaranteed to have a
higher resonant frequency. The relaxation via this route can be
estimated for a given detuning
$\delta=\hbar\omega-(\sqrt{\Delta^2+J^2}-J)$ between the cavity and
the qubit singlet similarly to the Purcell effect discussed in
Ref.~\onlinecite{houck}. The presence of the cavity modifies the
Hamiltonian by two terms, $H_{\rm
cav}=(\hbar\omega+1/2)\hat{a}^\dagger\hat{a}$ and $H_{\rm
cav-q}=\gamma(\sigma_x^{(2)}-\sigma_x^{(1)})(\hat{a}+\hat{a}^\dagger)$.
The first excited state corresponding to $|\tilde{\phi}_a\rangle$
has a photonic nature with approximately
$p=2(b-a)^2\gamma^2/\delta^2$ probability.
% in the d & \phi_a subspace the coupling energy to cavity is modified to $g\sqrt{2}(b-a)$ as is seen by simply evaluating the matric elements
% cf. standard Purcell effect
Here the coupling energy $\gamma=M_{\rm cav-q}I_{\rm p}I_{\rm rms}$
between the cavity mode and the qubits depends on the mutual
inductance $M_{\rm cav-q}$ between each qubit loop and the center
conductor (sign difference is built in the antisymmetric coupling)
and the rms current $I_{\rm rms}$  in the ground state of the
cavity. The relaxtion rate of the antisymmetric singlet limited by
the cavity quality factor $Q$ is thus simply $\Gamma_Q=p\omega/Q$.
If e.g. $\Delta/h=6$ GHz, $J/h=1$ GHz, $g/h=0.08$ GHz,
$\omega/2\pi=10$ GHz and $Q=10^4$ we get $1/\Gamma_Q=260$ $\mu$s.
This is long enough to detect the difference between the life times
of the subradiant and the superradiant states. Furthermore, a
numerical calculation for these values shows that the resonant
frequency of the cavity will be shifted down by about 1 MHz when the
singlet is excited compared to when the qubit is in the ground
state. This shift revealing the qubits' state is well detectable in
a microwave transmission measurement using a low-noise cold
amplifier in the same way as in Ref.~\onlinecite{wallraff} since the
width of the resonator transmission peak is comparable, i.e.
$\omega/(2\pi Q)=1$ MHz.

Owing to symmetry the effect of any global fluctuation is minimized
in the present system. Testing whether a significant part of the
relaxation is due to global fluctuations amounts to measuring the
lifetime of the state $|\phi_a\rangle$. Whether the result will be
positive or negative is not known but in any case this should give
valuable information about the origin of the noise.

We studied relaxation in an interacting two-qubit system exposed to
a global relaxation mechanism and showed how interference effects
lead to a dramatic deviation from the independent relaxation
picture. The small detuning of bath couplings leads to a slow
relaxation of the subradiant state while superradiant state decays
much faster than individual excitation. Experimental realization of
phenomena was discussed in detail in context of superconducting flux
qubits, where the phenomenon can be utilized to extract information
of an incompletely understood relaxation process and possibly
construct long-lived quantum states.

\end{document}